# Zero bias conductance peak in InAs nanowire coupled to superconducting electrodes

Nam-Hee Kim[1,§], Yun-Sok Shin[2,§], Hong-Seok Kim[1], Jin-Dong Song[3], Yong-Joo Doh[1*]

[1]Department of Physics and Photon Science, Gwangju Institute of Science and Technology, Gwangju 61005, Republic of Korea
[2]Sejong Academy of Science and Arts, Sejong 30099, Republic of Korea
[3]Korea Institute of Science and Technology, Seoul 02792, Republic of Korea

[*]E-mail: yjdoh@gist.ac.kr

[§]These authors contributed equally to this work.

**Abstract**

We report the occurrence of the zero-bias conductance peak (ZBCP) in an InAs nanowire coupled to PbIn superconductors with varying temperature, bias voltage, and magnetic field. The ZBCP is suppressed with increasing temperature and bias voltage above the Thouless energy of the nanowire. Applying a magnetic field also diminishes the ZBCP when the resultant magnetic flux reaches the magnetic flux quantum $h/2e$. Our observations are consistent with theoretical expectations of reflectionless tunneling, in which the phase coherence between an electron and its Andreev-reflected hole induces the ZBCP as long as time-reversal symmetry is preserved.

Nano-hybrid superconducting devices, which consist of low-dimensional nanostructures in contact with superconducting electrodes, provide a useful platform for developing novel quantum information devices. In particular, InAs nanowires (NWs) have attracted intense interests as a good material system for fabricating nano-hybrid superconducting devices owing to their outstanding electrical properties, such as high electron mobility, low effective mass, and relatively easy formation of low ohmic contacts. So far, a supercurrent transistor [1], NW quantum dot [2], quantum electron pump [3], Cooper pair splitter [4], and gate tunable superconducting qubit [5, 6] have been developed with InAs NWs. As a signature of a Majorana fermion, which is a particle identical to its own antiparticle, the zero-bias conductance peak (ZBCP) has also been observed in an InAs NW coupled to a superconducting electrode [7]. Because the ZBCP can also be caused by reflectionless tunneling [8] due to an incomplete superconducting proximity effect, more extensive studies are required to verify the physical origin of the ZBCP observed in InAs NWs.

The superconducting proximity effect in nano-hybrid superconducting junctions can be understood with the Andreev reflection [9], where an electron incident upon the interface between a normal metal (N) and superconductor (S) can be retro-reflected into the N region as a phase-conjugated hole while leaving a Cooper pair in the S region. As a result, the Andreev reflection occurring at an ideal N–S interface results in a twofold increase of the conductance for bias voltages smaller than $\Delta/e$, where $\Delta$ is the superconducting energy gap. With a non-ideal N–S interface, there exists a finite probability for the incident electron to be partially reflected as an electron and also retro-reflected as a hole via the Andreev reflection. When the mirror-reflected electron undergoes several scatterings with disorder centers in the N region and is incident on the N–S interface again, an additional Andreev reflection can occur. This results in another retro-reflected hole tracing back the previous scattering paths and continuing successive Andreev reflections. This process, which is called reflectionless tunneling [8], can also cause a ZBCP in a sample including the N–S interface. So far, the ZBCP due to reflectionless tunneling has been observed in several nano-hybrid superconducting junctions made of graphene [10] and InAs NW [11]. However, the observed values of the characteristic voltage and temperature for the ZBCP are not consistent with the theory [8], which has been attributed to a degradation effect at the N–S interface reducing the diffusion constant [11]. Here, we report the observation of a ZBCP obtained from an InAs NW in contact with superconducting PbIn electrodes. Our extensive studies of ZBCPs with a varying bias voltage, temperature, and magnetic field revealed that our observed ZBCP is due to the phase-coherent quantum electronic transport, which is consistent with the reflection tunneling theory.

InAs NWs were grown on a Si substrate by ultrahigh vacuum molecular beam epitaxy with gold particles (~50 nm diameter) as the catalyst [12]. The as-grown InAs NWs were transferred to a highly *p*-doped Si substrate covered by a 300 nm thick oxide layer. Source and drain electrodes were patterned by conventional *e*-beam lithography, followed by *e*-beam evaporation of 170 nm thick PbIn film. Prior to the metal deposition, the NW surface was cleaned with an Ar ion beam to remove a presumed native oxide layer. The PbIn alloy source was made by using a mixture of lead (Pb) and indium (In) pellets with a weight ratio of 9:1. The 10 nm thick Au film was used as a capping layer. Figure 1a shows a scanning electron microscopy (SEM) image of sample **D1** after the device fabrication process was completed. The superconducting transition temperature and perpendicular critical field of the PbIn film were obtained as $T_C$ = 7.3 K and $B_C$ = 0.24 T, respectively [13]. The differential resistance d$I$/d$V$ was measured by using a standard lock-in technique of superimposing a small AC signal with a frequency of 37.77 Hz onto the DC bias current. For low-noise measurement, two-stage resistor–capacitor (RC) filters (cutoff frequency = 10 kHz) and π filters were connected in series to the measurement leads [14].

Figure 1b displays the conductance $G$ of sample **D2** as a function of the backgate voltage $V_g$ at $T = 10$ K. This indicates $n$-type conduction of the InAs NW. The electron mobility was $\mu = dG/dV_g \times L^2/C_g = 3800$ cm$^2$/Vs, where $L$ is the channel length and $C_g$ is the gate capacitance between InAs NW and the Si substrate [15]. The diameter and channel length of sample **D1** were $d = 80$ nm and $L = 220$ nm, respectively. Those for sample **D2** were $d = 60$ nm and $L = 590$ nm, respectively. For sample **D2**, the carrier density is about $n \sim 2.8 \times 10^{17}$ cm$^{-3}$ and the mean free path is $l_m = 41$ nm at $V_g = 0$ V.

When the temperature was lowered below $T_C$, the sample resistance started to increase and display an insulating $R(T)$ behavior (see Figure 2a). At lower temperatures near $T^* = 3.6$ K, the insulating behavior changed to the metallic one. The former insulating $R(T)$ curve near $T_C$ can be understood with the Blonder–Tinkham–Klapwijk (BTK) theory, which considers the interfacial barrier effect at the N–S contact. According to BTK theory, the current flowing through the N–S interface is given by [16]

$$I_{NS} = \frac{1+Z^2}{eR_N} \int_{-\infty}^{\infty} [f_0(E - eV) - f_0(E)] \times [1 + A(E) - B(E)]dE, \qquad (1)$$

where $Z$ is a parameter characterizing the height of a presumable tunnel barrier at the N–S interface ($Z = 0$ for ideal transparency), $V$ is the bias voltage, $R_N$ is the normal-state resistance, $f_0$ is the Fermi–Dirac distribution function, $A(E)$ is the probability of the Andreev reflection, and $B(E)$ is the probability of the specular reflection. For $E < \Delta$, $A(E)$ is given by $\Delta^2/[E^2 + (\Delta^2 - E^2)(1 + 2Z^2)^2]$, and $B(E)$ is $1 - A(E)$. The $R(T)$ curve near $T_C$ fitted well with the calculation result of $(dI_{NS}/dV)^{-1}$ using $Z = 2$ and $\Delta_0 = 1.14$ meV as fitting parameters, as shown in Figure 2a. Here, $\Delta_0$ is a superconducting gap energy of the PbIn electrode at zero temperature. The deviation between the calculation and experimental data occurring at lower temperatures is discussed later.

The differential conductance $G = dI/dV$ was plotted as a function of $V$ at the base temperature $T = 2.4$ K (see Figure 2b). The $G(V)$ curve exhibited a ZBCP and additional peak structure at $V_{gap} = 2.3$ meV. The overall suppression of $G$ below $V_{gap}$ was attributed to the formation of a non-ideal N–S interface, and the $dI/dV$ peak position on the $V$ axis corresponds to the superconducting gap voltage $\Delta/e$. Because there existed two N–S interfaces coupled in series via the InAs NW, the $V_{gap}$ value resulted in $\Delta_{PbIn}(2.4\ K) = 1.15$ meV, where $\Delta_{PbIn}$ is the superconducting gap energy of PbIn electrodes. This result is consistent with the fitting result of $R(T)$ curve and other reports [14, 17].

We note that the ZBCP diminished at a characteristic voltage of $V^* = 0.22$ mV. Increasing the lock-in bias voltage showed a similar suppression of the ZBCP, as shown in Figure 2c. Because reflectionless tunneling arises from the quantum phase conjugation between the incident electron and Andreev-reflected hole, which traces back the path of the scattered electron, the characteristic voltage $V^*$ for observing the ZBCP can be related to the Thouless energy $E_{Th} = eV^* = 0.22$ meV of the InAs NW. Then, the phase coherence time is given by $\tau_\phi = \hbar/E_{Th} = 3.0$ ps, which results in the phase-coherence length of $L_\phi = \sqrt{D\tau_\phi}$ = 205 nm, where $\hbar = h/2\pi$ is the reduced Planck's constant, $D = v_F l_m/3 = 140$ cm$^2$/s is the diffusion coefficient, and $v_F = 1.04 \times 10^6$ m/s is the Fermi velocity [18]. Because $L_\phi$ is close to the channel length $L = 220$ nm, we concluded that the whole NW segment between two PbIn electrodes contributes to the phase-coherent reflection tunneling to result in the observed ZBCP feature.

As we increased the temperature, the ZBCP height gradually decreased and disappeared at the characteristic temperature $T^* = 3.6$ K, as shown in Figure 2d. This temperature-dependent behavior of the ZBCP is consistent with the metallic $R(T)$ curve below $T^*$ in Figure 2a. Because the thermal coherence length [19] is given by $\xi_T(T) = \sqrt{\hbar D/k_B T}$, we obtained $\xi_T(T^*) = 170$ nm, which is comparable to $L_\phi$. Thus, the suppression of ZBCP at higher temperatures than $T^*$ was attributed to a dephasing effect due to thermal fluctuations, resulting in $\xi_T(T) < L_\phi$. We note that $T^*$ is higher than the dephasing temperature estimated from the Thouless energy: $T_{Th} = eV^*/k_B = 2.6$ K. This underestimation of $V^*$ can be caused by a finite background conductance at voltages below $V_{gap}$ on the $dI/dV$ vs. $V$ curve resulting from a finite interfacial barrier at the N–S interfaces of the InAs NW sample.

When the magnetic field $B$ was applied perpendicular to the substrate, the ZBCP was suppressed and eventually vanished at $B^* = 36$ mT, as shown in Figure 3a and 3b. Because the critical magnetic field $B_C$ of the superconducting PbIn electrode [13] was found to be larger than 200 mT, the disappearance of the ZBCP is not caused by the magnetic-field-induced suppression of superconductivity in the PbIn electrode. Rather, it was attributed to broken time-reversal symmetry, which causes a phase shift between the electron and its retro-reflected hole during the reflectionless tunneling process. Because the total phase shift is given by [8] $\Delta\phi = 2\pi BA/\Phi_0$, where $\Phi_0 = h/2e$ is the magnetic flux quantum and $A$ is the effective area enclosed by the superconductor and a loop formed by the trajectory of the scattered electron and retro-reflected hole, we obtained $A = 0.062$ μm$^2$ from $\Delta\phi(B^*) = 2\pi$. When the lateral width is approximated by the NW diameter, the effective length turned out to be $L_{eff} = 0.78$ μm, which is much longer than the channel length. The difference between $L$ and $L_{eff}$ can be explained by the London penetration depth ($\lambda$) of PbIn, which resulted in $\lambda = L_{eff} - L = 560$ nm. This value is comparable to $\lambda \sim 450$ nm in the previous report [14].

In conclusion, we observed ZBCP in an InAs NW contacting PbIn superconducting electrodes. The temperature, bias voltage, and magnetic field dependencies of the ZBCP were consistent with the theoretical expectations of the phase-coherent electronic transport via reflectionless tunneling. The characteristic energy for observing the ZBCP was determined by the Thouless energy of the NW, while the phase coherence between the electron and its retro-reflected hole was preserved. Our experimental studies would provide clear criteria for discriminating between ZBCPs due to the Andreev reflection and due to Majorana fermions.

## Acknowledgment

This work was supported by the "GIST-Caltech Research Collaboration Grant" funded by the GIST in 2017.

## Figure Captions

**Figure 1.** (a) SEM image of the InAs NW device (**D1**) with the typical measurement configuration. The current bias $I$ is applied between two PbIn electrodes (+I and –I) through the InAs NW, and the voltage $V$ is monitored between the +V and –V electrodes. (b) Differential conductance $G$ of sample **D2** with the back-gate voltage $V_g$ at $T = 10$ K. $V_g$ was applied to the Si substrate, and the source-drain bias voltage was $V_{SD} = 1$ mV.

**Figure 2.** (a) $R$ vs. $T$ plot obtained from sample **D1**. The superconducting transition temperature was $T_C = 7.3$ K. The red solid line represents the fitting result (see text). (b) Normalized conductance ($G$) as a function of $V$ at $T = 2.4$ K. The characteristic voltage $V^*$ for the ZBCP was about 0.22 mV, and the gap voltage $V_{gap}$ was 2.3 mV. (c) Lock-in voltage ($V_{ac}$)

dependence of the ZBCP at $T = 2.4$ K. From top to bottom, $V_{ac} = 3.9$ µV (black), 11 µV (red), 22 µV (blue), 130 µV (green), and 310 µV (magenta). The curves are offset for clarity. (d) Temperature dependence of the ZBCP. From top to bottom, $T = 2.4$ K (black), 2.7 K (red), 3.0 K (blue), 3.3 K (green), and 3.6 K (magenta). The curves are offset for clarity.

**Figure 3.** (a) $G$ vs. $V$ curve with varying magnetic field at $T = 2.4$ K. From top to bottom, $B = 0$ mT (black), 14 mT (red), 25 mT (blue), 29 mT (green), and 34 mT (magenta). The curves are offset for clarity. (b) Zero-bias conductance as a function of the $B$ field. $B^*$ was about 36 mT, and $B_C$ was about 280 mT.

**Figure 1**

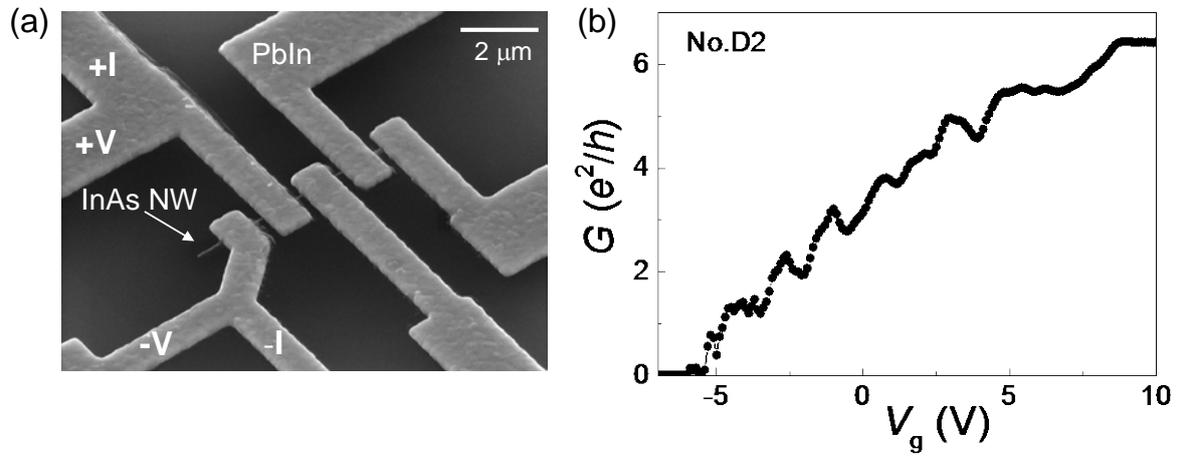

**Figure 2**

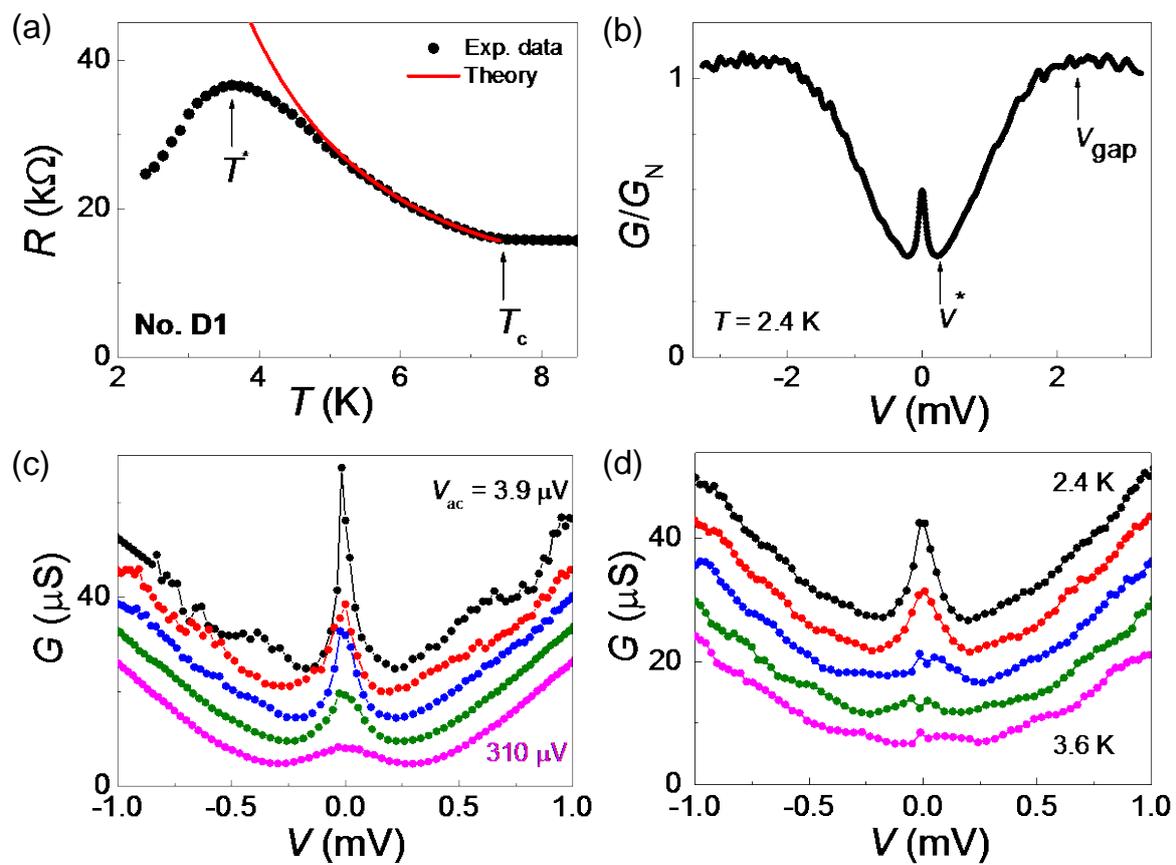

**Figure 3**

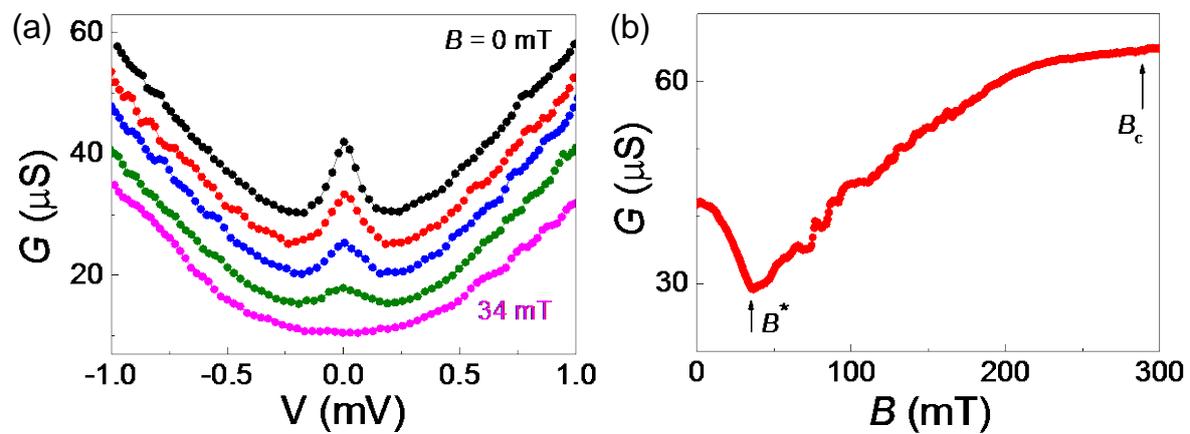